# Anomalous and Topological Hall effect in Cu doped $Sb_2Te_3$ Topological Insulator

Abhishek Singh[1, a], Vinod K Gangwar[1,a], Prashant Shahi[2], Debarati Pal[1], Rahul Singh[1], A. K. Ghosh[3], Swapnil Patil[1], Jinguang Cheng[2],and Sandip Chatterjee[1,*]

[1]Department of Physics, Indian Institute of Technology (Banaras Hindu University), Varanasi-221005, India

[2]Beijing National Laboratory for Condensed Matter Physics and Institute of Physics, Chinese Academy of Sciences, Beijing 100190, China

[3]Department of Physics, Banaras Hindu University, Varanasi-221005, India

## Abstract

The magneto-transport and magnetization measurements of $Sb_{1.90}Cu_{0.10}Te_3$ were performed at different temperatures and different fields. Magneto-transport measurement at high field indicates the coexistence of both bulk and surface states. The magnetization shows the induced antiferromagnetic ordering with Cu doping and the observed quantum oscillation in it indicates that magnetization in $Sb_{1.90}Cu_{0.10}Te_3$ is the bulk property. The non linearity in Hall data suggests the existence of anomalous and topological Hall effect. The anomalous and topological Hall effect (THE) from measured hall data of Cu doped $Sb_2Te_3$ topological insulator have been evaluated.

Finding anomalous [1] and topological Hall effect [2, 3] in topological states of matter is highly interesting. Generally, topological Hall effect is observed in B20 phase materials [4, 5] but the same has also been observed in some other kind of materials [6-8]. While this phenomenon can be understood in ferromagnetic materials, but recently this has also been reported in antiferromagnetic system [9]. Here we have presented the existence of topological Hall effect (THE) in Cu doped $Sb_2Te_3$ topological insulator. In topologically non-trivial HgTe [10, 11] the THE has already been predicted. However, in antiferromagnetic GdPtBi [9], whose electronic structure is similar to the HgTe, the existence of THE has been reported. We have shown the signature of THE from the Hall effect measurement. Magnetic data show the anti-ferromagnetic ordering. The THE observed in the present composition even at room temperature is highly interesting from the application point of view.

Moreover, recently, topological insulators (TIs) have attracted a great attention to the scientific community for their novel physical properties. The surface state in TIs which is conducting unlike insulating bulk stateis topologically protected by time reversal symmetry (TRS) due to the locking of spin and orbital states [12]. The delocalized Topological surface states (TSS) remain unaffected with nonmagnetic doping and their excitation spectrum within the bulk energy gap exhibits the characteristic Dirac dispersion because of this time reversal symmetry. Furthermore, the Dirac character of the surface carriers can be shown by evaluating the temperature and field dependence of the SdH oscillations and by extracting the Berry phase.

On the other hand, in systems where time-reversal symmetry (TRS) is spontaneously broken, there are contributions to transverse velocity from both extrinsic effects due to spindependent scatteringand intrinsic effects related to real space and momentum space Berry phase mechanisms [1-2, 13]. The former isrelevant in systems with non-coplanar spin textures with finite scalar spin chirality while the latter generically occurs in TRS-broken systems originating from the spin-orbit interaction-induced Berry curvature of the filled bands. The anomalous Hall effect (AHE) due to magnetic texture is most often associated with finite spin chirality and tends to exhibit relatively small anomalous Hall angles ~ 0.01 (such as $SrFeO_3$ [14] or$Pr_2Ir_2O_7$ [15]).

But in ferromagnetic systems intrinsic band-structure-based effects are usual and significantly large. However, it has been suggested theoretically that effects due to both magnetic texture and strong spin-orbit coupling may exist in non-collinear antiferromagnets which lead to significant Hall responses [16]. In the present study we have shown the existence THE in Cu doped $Sb_2Te_3$ topological insulator.

The focus of the present study is the interplay of the magnetic structure and strong spin-orbit coupling. A valuable perspective of the coupling of the conduction electrons to the magnetic ordering can be obtained from magneto-transport. In the present investigation, from the magneto-transport and magnetic measurements we have shown the existence of surface and bulk states and also we see here that for the magnetic ordering bulk state plays the main role. The variation of resistivity as a function of temperature of $Sb_{1.90}Cu_{0.10}Te_3$ has been measured [Supplementary Materials]. It is observed that resistivity value increases with increasing magnetic field. $Sb_{1.90}Cu_{0.10}Te_3$ shows a large linear MR (nearly 45%) at low temperature but with increase of temperature MR decreases. For a high magnetic field, Landau-level induced SdH oscillations were observed at low temperatures [Supplementary]. The Landau level fan diagram obtained from quantum oscillation shows an intercept at ∼-0.479 (fig.1 (a)), indicating that the Dirac fermions dominate the transport properties due to the additional Berry phase π.

The variations of magnetization as a function of magnetic fieldas well as temperature are shown in fig.2. In fig.2, we observe that magnetization increases with increasing magnetic field. The magnetic susceptibility ($\chi$) vs. T curve (inset of fig.2), has been fitted using the following equation

$$\chi = \chi_0 + (D \times T) + C/(T-\theta) \tag{1}$$

First two terms in equation (1) describe the diamagnetic contribution in magnetic susceptibility whereas third term is indicating about the Curie-Weiss expression. Best fit of experimental data gives θ = -0.46K, the negative value of θ clearly indicates the anti-ferromagnetic nature.

For further confirmation of the anti-ferromagnetism, M-H measurement is carried out at different temperatures (*viz.* at 2K, 5K, 10K, 25K, 50K, 200K and 300K) under the applied field range of -5T to 5T. It is clear from the M-H curve that M increases with the increase in H. Moreover, no saturation is seen even at the applied field of ±5T, which is a clear signature of anti-ferromagnetic ordering. Interestingly, nature of curve remains anti-ferromagnetic even at room temperature (i.e. 300K). Hence it is clear that doping of Cu tuning the nature of $Sb_2Te_3$ from diamagnetic to anti-ferromagnetic as pure $Sb_2Te_3$ is a well established diamagnetic material [18]. Moreover, for a higher magnetic field, Landau-level induced de-Has van Alphen (dHvA) oscillations are observed at low temperatures (Supplementary). The obtained intercept is ~0.074 (close to zero) indicating the absence of time reversal symmetry in bulk state (fig.1.(b)).

Figure 3 shows the magnetic field variation of Hall resistivity at different temperatures for $Sb_{1.90}Cu_{0.10}Te_3$. We can easily observe three features from $\rho_{xy}$ vs. B graph (i) background is linear at large field (ii) it is non linear at low field (iii) it switches its sign from positive to negative (except at 50K), before reaching at zero field. Linearity in the graph clearly indicates the ordinary Hall Effect (OHE) whereas presence of multicarrier and anomalous Hall effect (AHE) may be two responsible factors for the non-linearity in the Hall graph [shown in Fig.3].

We have tried to fit the data by two band model but the extracted parameters were not feasible (not shown here), therefore, multicarrier contribution can be excluded easily. Hence, one can conclude that AHE is the responsible factor for the nonlinearity in the graph at low field. Presence of AHE is also an indication of magnetic ordering in sample [19]. Moreover, Hall resistivity is switching sign from positive to negative at low field before reaching the zero value [shown in Inset of fig.3], indicating the presence of topological Hall Effect (THE). Hence, total Hall resistivity may be expressed as the combination of three terms OHE, AHE and THE

$$\rho_{xy} = R_o B + R_s M + \rho^{TH} \qquad (2)$$

Here $\rho_{xy}$ shows the total Hall resistivity. First term in right hand side represents the ordinary Hall effect (OHE) where $R_0$ is the ordinary Hall Effect coefficient. Second term indicates the presence of anomalous Hall Effect (AHE), $R_s$ is the anomalous Hall coefficient and M is the out of plane magnetization. Third term represents the topological Hall Effect.

The coefficient of Anomalous Hall Effect (AHE) can be represented as

$$R_s = a\rho_{xx} + b\rho_{xx}^2 \qquad (3)$$

The First linear term $a\rho_{xx}$ indicates the skew scattering term whereas the second quadratic term $b\rho_{xx}^2$ represents the scattering independent contribution in Hall resistivity. Tian et al. claimed that the linear term can be neglected in the bulk [20], moreover, it is noteworthy that scattering at the surface of the topological insulator is very less significant. Porter et al. have reported that the screw scattering term is only significant in nearly perfect crystals, taking account of all these factors $\rho_{xx}$ can be neglected [21]. Hence it is clear from the above discussion that the anomalous Hall Effect is mainly dominated by quadratic ($b\rho_{xx}^2$) term. Topological Hall effect is absent ($\rho^{TH}$

= 0) at high magnetic field. Although THE is an indication of non-coplanar spin texture but it is still an indirect method for the detection of magnetic skyrmions as compared to direct method by magnetic force microscope and Lorentz transmission electron spectroscopy. All these methods are surface sensitive only and the surface of $Sb_{0.90}Cu_{0.10}Te_3$ is non magnetic (which will be discussed later) hence one cannot use these techniques to get information about the skyrmions, hence at the moment, THE may be the best possible techniques to detect the type of topologically nontrivial structure such as magnetic skyrmions [22].

It is observed that at higher field when $\rho^{TH} = 0$, Equation (2) can be simplified as

$$\frac{\rho_{xy}}{B} = R_o + \frac{b\rho_{xx}^2 M}{B} \qquad (2)$$

If we linearly fit the $\frac{\rho_{xy}}{B}$ vs $\frac{b\rho_{xx}^2 M}{B}$ graph at the high magnetic field we can extract $R_0$ and b from the intercept and the slope of the curve respectively. We obtained a very good fit at high magnetic field with the fitting parameters $R_0$ and b as shown in Fig 4. Subtracting ordinary Hall resistivity ($R_o B$) and anomalous Hall resistivity ($b\rho_{xx}^2 M$) from the total Hall resistivity ($\rho_{xy}$) data in the full range of magnetic field, topological Hall resistivity ($\rho^{TH}$) has been extracted. In fact, the signature of the existence of THE is observed from the very low temperature (2K) and gradually it decreases and at 50K it is almost zero and with further increase of temperature the value further increases and finally at 300K it shows maximum value at minimum field (~0.25T).

Gallagher et al. and Yu et al.have shown that THE increases with increasing temperature and thermal fluctuation are necessary to stabilize skyrmions [5, 23]. The maximum reported temperature at which THE observed was 275K [23]. This is the first report showing THE at room temperature in the present investigation to the best of our knowledge. Moreover, the obtained topological Hall Effectis as large as 1200 nΩ-cm which is even higher than the earlier reported values (8,22,24). The highest reported value for THE was 1800 nΩ–cm which was obtained at very low temperature i.e. 2.5K [9]. To investigate the origin of anti-ferromagnetism and to collect the information about the valence state and chemical bonding of the elements present in the lattice, we have used X-ray photoemission spectroscopy (Supplementary). It is observed from the analysis of the data that Cu is in $Cu^{2+}$ state. Presence of $Cu^{2+}$ spin state might be a reason for the presence of antiferomagnetism in our sample.

In fact spin texture is the origin of THE. Linear fit of MR vs. $M^2$ (shown in Fig.4(i)) data is clearly indicating the presence of spin texture in our sample which is also consistent with M(H) and M(T) analysis showing magnetic ordering. It is also clear from XPS analysis that Cu is in +2 states i.e. $3d^9$ states (Supplementary), hence it should contain one unpaired electron and the corresponding magnetic moment for this state should be √3 Bohr magneton. The spins are anti-parallel aligned which gives rise to antiferromagnetic state as well as skyrmion phase in the sample. When electrons travel through the skyrmion it is deviated which gives rise to THE. Moreover, from the above discussion it is clear even with the Cu doping the strong spin-orbit coupling exists in this Cu doped $Sb_2Te_3$. Also we have seen that time reversal symmetry is broken at the bulk state. Therefore, bulk state is the origin of this THE. This is also supported by the magnetic data. To further establish the bulk origin of THE we have also carried out the band structutre calculation by DFT (shown in Fig.5).

The most important role of magnetic texture to the Topological Hall effect (THE) is that of skyrmionic spin textures [25, 26]. Recently, in anti-ferromagnets, theoretically it is observed that a band structure induced THE might be due to the presence of non-collinear spin structure which breaks both the time reversal and lattice symmetries [16, 27]. Such a situation is observed in the present system as the AFM spinscant at finite magnetic field. Magnetic datasupport both the AFM ordering and a FM canting as is evident by an increase in magnetization with field and existence of hysteresis loop. As a matter of fact, in an AFM system with such broken time-reversal symmetry and spin-orbit coupling the THE is predicted to be significant.

Furthermore, in order to investigate the origin of the THE observed in Cu-doped $Sb_2Te_3$ at finite magnetic field the effect of the spin texture on the electronic structure of this Cu-doped $Sb_2Te_3$ is considered, which might be the source of THE as is observed in FM system [7, 8, 26], if driven to anti-crossings near Fermi level. We have also calculated the electronic structure of undoped $Sb_2Te_3$ where no anti crossing is observed. But when Cu is doped anti-crossing at Fermi level is observed and with application of magnetic field avoided crossings developed. This may carry the significant Berry curvature. The combination of large spin-orbit coupling and critical band alignment of the system amplifies the physical consequence of the Cu spin texture on these bands. Therefore, significant Berry phase contributes to the THE due to the development of spin-texture with the evolution of electronic structure.

In conclusion, Hall data indicate the existence of Anomalous and Topological Hall effects even at room temperature. Magnetization data indicates that the Cu doped $Sb_2Te_3$ is in the AFM state. But with application of magnetic field the spin structure cants ferromagnetically which is observed from the appearance of hysteresis loop in M(H) curve. The observed THE might be due to the spin textures which is clear from the linear variation of MR with $M^2$. The band structure confirms the existence of spin texture. we have measured the magneto-transport and magnetization at different temperatures and different fields. We find the coexistence of both bulk and surface states from the quantum oscillations (Supplementary file). The observed room temperature anti-ferromagnetism is due to the $Cu^{2+}$ spin state of Cu. Moreover, THE has been established at room temperature which may very promising spintronics devices working at room temperature.

**Materials Method**

The single crystal of $Sb_{1.90}Cu_{0.10}Te_3$ was grown by using modified Bridgman method [17], the single crystal of $Sb_{1.90}Cu_{0.10}Te_3$ was grown by melting stoichiometric mixture of constituent elements. The mixture of high purity (99.99%) Sb,Te and Cu were sealed in evacuated quartz ample and heated upto 900 ˚C by raising the temperature at 200˚C per hour and was kept at that temperature for 10 hours and then it was slowly cooled to 550˚C at the rate 5˚C per hour and after that it was naturally cooled to room temperature. Thus the obtained crystal was easily cleaved along (*00l*) direction. The electrical transport properties were carried out by using physical property measurement system (PPMS, Quantum Design). The magnetic properties were carried out by using magnetic property measurement system (MPMS, Quantum Design).

**References:**

[a] Both the authors have equal contribution.

* Corresponding author e-mail id: schatterji.app@iitbhu.ac.in

**Figure Captions:**

Fig.1.(a) SdH oscillations from the longitudinal resistance and Landau level indexing (Fan Diagram) with inverse magnetic field and linearly fitted curve (red line). (b)dHvA oscillations from the magnetization and Landau level indexing (Fan Diagram) with inverse magnetic field and linearly fitted curve(red line).

Fig.2. Field dependence of magnetization of $Sb_{1.90}Cu_{0.10}Te_3$ at different temperatures. The inset represents Temperature dependence of magnetization in ZFC mode at an applied magnetic field of 1000 Oe.

Fig.3. Magnetic field dependence of the Hall resistivity of $Sb_{1.90}Cu_{0.10}Te_3$ at different temperatures. Inset shows closed view of hall data at low field.

Fig.4. Hall resistivity at temperatures 2K, 5K, 10K, 25K, 50K, 200K, 300K in (a),(b),(c) (d),(e),(f) and (g) respectively. The Solid lines are the fittedcurves ofHall resistivity ($\rho_{xy}$) using the relation that $\rho_{xy} = R_0B + b\rho_{xx}^2 M$ withthe fitting parameters $R_0$ and b. (h) Topological Hall effect ($\rho^{TH}$) at different temperatures with respect to magnetic field.(i) Linear fit of MR vs. $M^2$.

**Fig.5.** Upper panel: Electronic structure of $Sb_2Te_3$. Middle panel: Electronic structure of Cu doped $Sb_2Te_3$ without magnetic field. Lower poanel: Electronic structure of Cu doped $Sb_2Te_3$ with magnetic field.

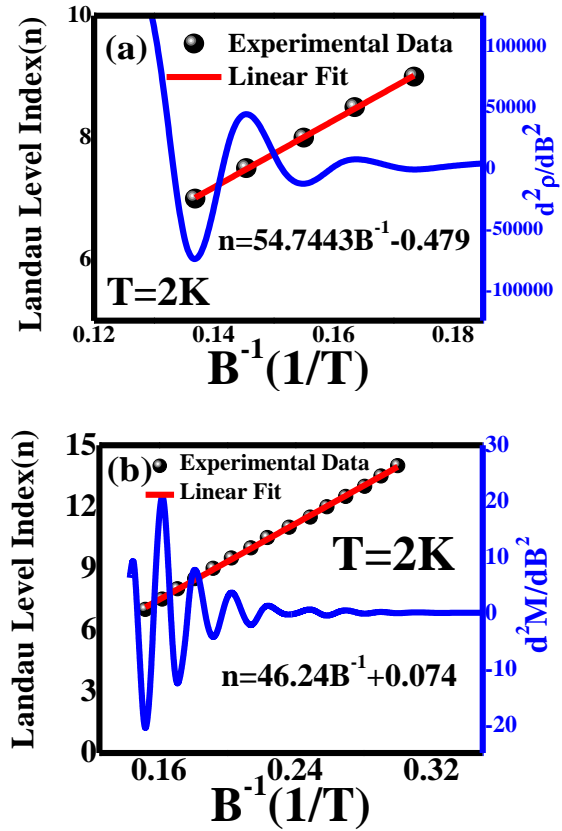

**Fig.1**

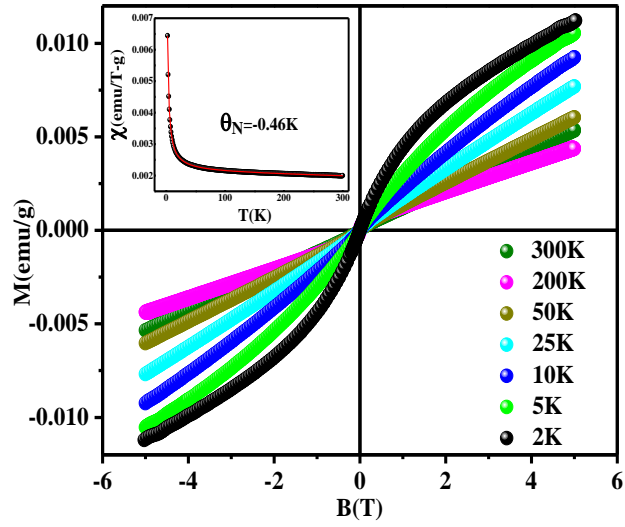

**Fig.2**

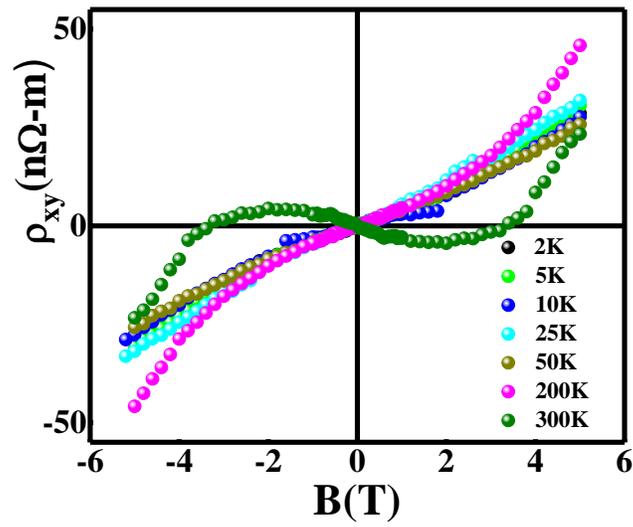

**Fig.3**

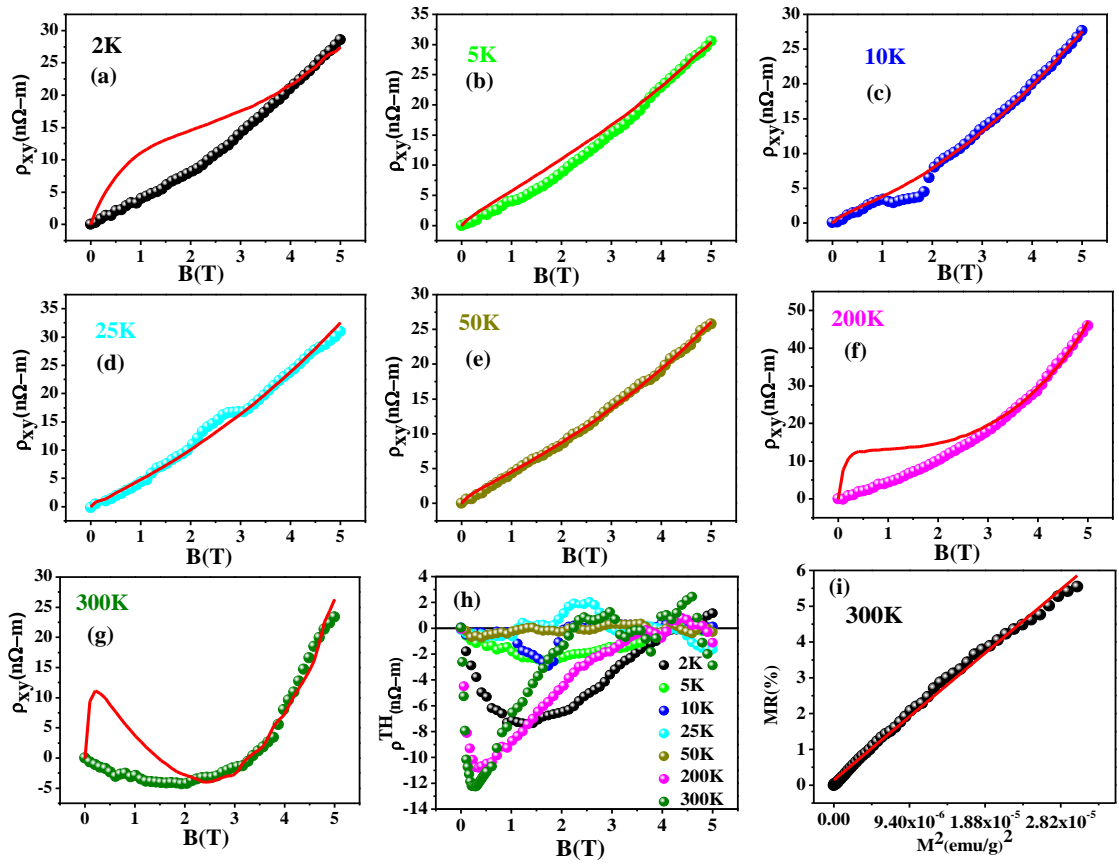

**Fig.4**

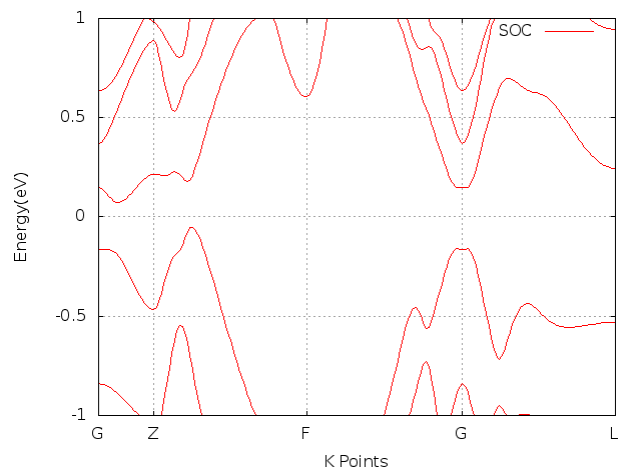

(a)

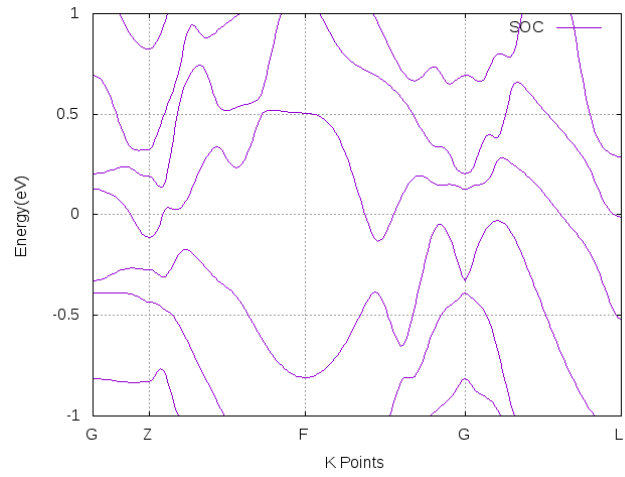

(b)

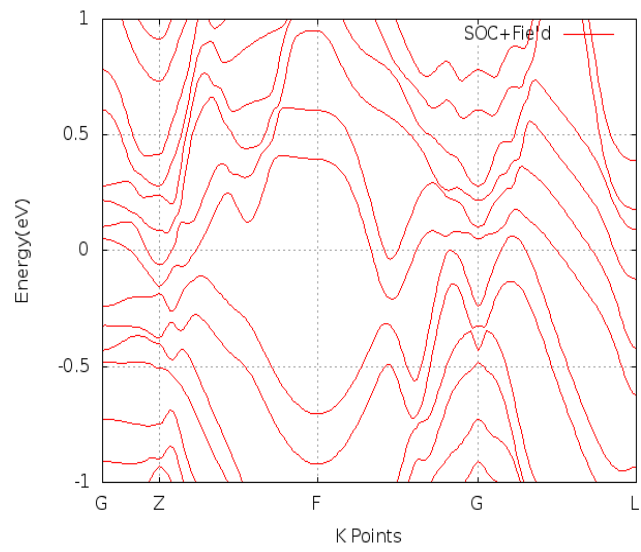

(c)

**Fig.5**